\documentclass[a4paper,12pt]{article}
\usepackage{graphicx,cite,epsfig}
\usepackage{a4}
\usepackage{latexsym, amssymb} 

\newcommand{\lapprox}{%
\mathrel{%
\setbox0=\hbox{$<$}
\raise0.6ex\copy0\kern-\wd0
\lower0.65ex\hbox{$\sim$}
}}
\newcommand{\gapprox}{%
\mathrel{%
\setbox0=\hbox{$>$}
\raise0.6ex\copy0\kern-\wd0
\lower0.65ex\hbox{$\sim$}
}}

\def\bt{\begin{table}}
\def\et{\end{table}}

\def\N0{\widetilde{\chi}^0}

\newcommand{\be}{\begin{equation}}
\newcommand{\ee}{\end{equation}}
\newcommand{\bea}{\begin{eqnarray}}
\newcommand{\eea}{\end{eqnarray}}

\def\thefootnote{\fnsymbol{footnote}}

\newcommand{\chione}{\widetilde{\chi}_1^\pm}
\newcommand{\chionezero}{\widetilde{\chi}_1^0}
\newcommand{\chitwo}{\widetilde{\chi}_2^0}

\begin{document}

\begin{titlepage}

\begin{flushright}
{\small December 6, 2007 \\ 
HRI-P-07-07-002 \\
HRI-RECAPP-07-08}
\end{flushright}

\vspace*{0.2cm}
\begin{center}
{\Large {\bf Distinguishing the Littlest Higgs model with T-parity\\[0.2cm] 
from supersymmetry at the LHC using trileptons}}\\[20mm]  
AseshKrishna Datta, Paramita Dey, Sudhir Kumar Gupta, \\ 
Biswarup Mukhopadhyaya and Andreas Nyf\/feler\footnote{E-mail:
  (asesh,paramita,guptask,biswarup,nyf\/feler)@mri.ernet.in}\\[0.5cm]  

{\it Regional Centre for Accelerator-based Particle Physics \\
Harish-Chandra Research Institute \\
Chhatnag Road, Jhusi
\\ Allahabad - 211 019, India}\\[20mm]
\end{center}

\begin{abstract}
We analyse hadronically quiet trilepton signatures in the T-parity conserving
Littlest Higgs model and in R-parity conserving supersymmetry at the Large
Hadron Collider. We identify the regions of the parameter space where such
signals can reveal the presence of these new physics models above the Standard
Model background and distinguish them from each other, even in a situation
when the mass spectrum of the Littlest Higgs model resembles the
supersymmetric pattern.
\end{abstract}

\pagestyle{plain}

\end{titlepage}

\setcounter{page}{1}

%------------------------------------------------------------------------

\renewcommand{\thefootnote}{\arabic{footnote}}
\setcounter{footnote}{0}

Little Higgs models have emerged recently as a class of alternative new
physics schemes for justifying the smallness of the Standard Model (SM) Higgs
boson mass~\cite{LH_original,LH_reviews}.  Precision
electroweak constraints normally imply that the mass scale of the new
particles in such theories has to be on the order of several
TeV~\cite{LH_EW_tests}, thus leading to a somewhat distasteful degree of
fine-tuning. The problem is circumvented through the coinage of an additional
discrete symmetry, the so-called T-parity~\cite{T_parity,LHT_Low}, whereby all
particles in the spectrum are classified as T-even(odd). This allows one to
have the Higgs mass protected from quadratic divergences, and at the same time
see a spectrum of additional gauge bosons, scalars and fermions, in the mass
range of a few hundred GeV's, with the lightest T-odd particle (LTP) being
stable.  Although it has been recently pointed out~\cite{Hill_Hill} that
T-parity can be broken by topological effects arising from anomaly terms, the
presence of such effects may not be absolutely certain in an ultraviolet
incomplete theory. Therefore, while the consequence of broken T-parity
requires careful attention~\cite{Hill_Hill,Barger_Keung_Gao}, it is desirable
from a phenomenological point of view to understand the collider signatures of
unbroken T-parity as well, together with its distinctive features in relation
to other new theories predicting weakly interacting massive stable particles.

The experimental signals of a scenario with T-parity have close resemblance
with those of supersymmetry (SUSY) with conserved R-parity. This is best
demonstrated in the context of a simple version of the model, namely, the
Littlest Higgs model with T-parity
(LHT)~\cite{LHT_Low,Hubisz_Meade,Chen_Tobe_Yuan}. With the LTP carrying away
substantial missing transverse momentum, accompanied by jets and/or leptons
rendered hard through the release of energy in the decay of heavy fermions,
the LHT, to a first approximation, may indeed look like SUSY. A number of
studies have already predicted interesting phenomenological consequences of
the LHT, for instance at the Large Hadron Collider
(LHC)~\cite{Hubisz_Meade,Chen_Tobe_Yuan,Freitas_Wyler,Belyaev_et_al,LHT_at_LHC,Cao_Chen}. A
few papers have also addressed the question of distinction between the LHT and
SUSY at the LHC~\cite{Chen_Cheung_Yuan,Nojiri_Takeuchi,LHT_SUSY_LHC} (for
non-collider signals, see Ref.~\cite{LHT_SUSY_non_collider}). However, there
still remains ample need for suggesting new ways of discriminating between the
two scenarios. Such a method is proposed in this note, where we emphasise the
role played by the non-strongly interacting sectors of such theories, and
analyse `hadronically quiet' trilepton signals at the LHC.

Of course, the new particles in SUSY and LHT, whose signals can mimic each
other, have different spins. However, extraction of spin information is
notoriously difficult at the LHC.  Another major difference between the LHT
scenario and SUSY lies in the non-existence of a counterpart for the gluino,
as the strongly interacting sector is untouched in Little Higgs models which
are essentially schemes for stabilisation of the electroweak sector. It has
been suggested in this light, for example, in \cite{Chen_Cheung_Yuan}, that in
the so-called `co-annihilation region', a SUSY scenario produces more
multi-jet events compared to the LHT, because of production channels involving
a gluino.  More importantly, since the gluino is a Majorana fermion, its
pair-production followed by cascade decays can produce like- and unlike-sign
charginos with equal probability, thus leading to a substantial enhancement of
final states with like-sign dileptons. While this can provide a tangible
discrimination between the two types of theories, it becomes unavailable if
the gluino is far too heavy (say, 2 TeV or more). In such cases, of course,
pair-production of same-sign squarks in SUSY and of same-sign heavy quarks in
LHT can give rise to like-sign dileptons. According to a recent
study~\cite{Nojiri_Takeuchi}, once one filters away the kinematical regions
where gluino cascades contribute, the ratio between like-and unlike-sign
dileptons is different in the two cases.  While such a claim is reassuring, it
is better to have some additional discriminating signals, preferably involving
the electroweak sector alone.  The purpose of this note is to suggest the
`hadronically quiet' trilepton channel in this context. It should also be
emphasised that we envision a situation where new signals are seen, and some
idea about the masses of the new particles is available from the hardness of
leptons and/or jets or the missing-$p_T$ distribution.

In the LHT a global symmetry $SU(5)$ is spontaneously broken down to $SO(5)$
at a scale $f \sim 1~\mbox{TeV}$. An $[SU(2) \times U(1)]^2$ gauge symmetry is
imposed, which is simultaneously broken at $f$ to the diagonal subgroup
$SU(2)_L \times U(1)_Y$, which is identified with the SM gauge group. This
leads to four heavy gauge bosons $W_H^\pm, Z_H$ and $A_H$ with masses $\sim f$
in addition to the SM gauge fields. The SM Higgs doublet is part of an
assortment of pseudo-Goldstone bosons which result from the spontaneous
breaking of the global symmetry. This symmetry protects the Higgs mass from
getting quadratic divergences at one loop, even in the presence of gauge and
Yukawa interactions. Electroweak symmetry is broken via the Coleman-Weinberg
mechanism and the Higgs mass is generated radiatively, leading naturally to a
light Higgs boson.  The multiplet of Goldstone bosons contains a heavy $SU(2)$
triplet scalar $\Phi$ as well. In contrast to SUSY, the new states which
cancel the quadratically divergent contributions to the Higgs mass due to the
top quark, gauge boson and Higgs boson loops, respectively, are heavy
fermions, additional gauge bosons and triplet Higgs states.

In order to comply with strong constraints from electroweak precision data on
the Littlest Higgs model~\cite{LH_EW_tests}, one imposes
T-parity~\cite{T_parity} which maps the two pairs of gauge groups $SU(2)_i
\times U(1)_i, i=1,2$ into each other, forcing the corresponding gauge
couplings to be equal, with $g_1 = g_2$ and $g_1^\prime = g_2^\prime$. All SM
particles, including the Higgs doublet, are even under T-parity, whereas the
four additional heavy gauge bosons and the Higgs triplet are T-odd. The top
quark has two heavy fermionic partners, $T_{+}$ (T-even) and $T_{-}$
(T-odd). For consistency of the model, one has to introduce the additional
heavy, T-odd vector-like fermions $u^i_H, d^i_H, e^i_H$ and $\nu^i_H$
($i=1,2,3$) for each SM quark and lepton field. For further details on the
LHT, we refer the reader to
Refs.~\cite{LHT_Low,Hubisz_Meade,Hubisz_et_al,Chen_Tobe_Yuan}. As shown in
Refs.~\cite{Hubisz_et_al,Asano_et_al,Hundi_Mukhopadhyaya_Nyffeler}, a scale
$f$ (which dictates the masses of most new particles) as low as
$500~\mbox{GeV}$ is compatible in the LHT with electroweak precision data.
Further constraints on the parameters of the LHT come from flavour
physics~\cite{constraints_flavor_physics}.

The masses of the heavy gauge bosons in the LHT are given by 
\be \label{gauge_boson_masses}
m_{W_H} = m_{Z_H} = g f \left(1 - {v^2 \over 8 f^2} \right) \approx 0.65 f,
\quad  
m_{A_H} = {f g^\prime \over \sqrt{5}} \left(1 - {5 v^2 \over 8 f^2} \right)  
\approx 0.16 f, 
\ee
where corrections of ${\cal O}(v^2/f^2)$ are neglected in the approximate
numerical values. Thus these particles have masses of several hundreds of GeV
for $f \sim 1~\mbox{TeV}$, although $A_H$, the heavy partner of the photon,
can be quite light, because of the small prefactor, and is usually assumed to
be the LTP. The masses of the heavy, T-odd fermions are determined by general
$3\times 3$ mass matrices in the (mirror) flavour space, $m_{q_H,l_H}^{ij}
\sim \kappa_{q,l}^{ij} f$ with $i,j=1,2,3$. We simplify our analysis by
assuming that $\kappa_q^{ij} = \kappa_q \delta^{ij}$. The parameter $\kappa_q
\sim {\cal O}(1)$ thus determines the masses of the heavy quarks in the
following way:
\be \label{mirror_quark_masses}
m_{u_H} = \sqrt{2} \kappa_q f \left(1 - {v^2 \over 8 f^2} \right), \quad 
m_{d_H} = \sqrt{2} \kappa_q f. 
\ee
Similarly, masses of the heavy leptons in the spectrum are determined by a
common parameter $\kappa_l$.  We further assume that the values for
$\kappa_{q,l}$ are not close to the upper bound $\kappa \leq 4.8$ (for $f =
1~\mbox{TeV}$) obtained from 4-fermion operators~\cite{Hubisz_et_al} and that
limits of $m > {\cal O}(100~\mbox{GeV})$ from direct searches at the Large
Electron Positron (LEP) collider apply to the mirror fermions in the LHT. Thus
our analysis takes $\kappa_{q,l}$ in the range $0.2 \lapprox \kappa_{q,l}
\lapprox 2$, thereby allowing all new heavy fermions to have masses ranging
from several hundreds of GeV to a TeV, for $f \sim 1~\mbox{TeV}$. For
our analysis we have used $\kappa_l = 0.4$, with $\kappa_q = 1$ and
$1.5$.\footnote{We also performed the same analysis for higher
$\kappa_l$ values ($\kappa_l=1$) and found the event rates both in the
LHT and SUSY to be too small to be observable after the cuts. We will
discuss this point later.} This yields masses of the heavy leptons and
quarks which are spaced relative to each other in a way often
encountered in SUSY for sleptons and squarks, so that the situation
where one spectrum fakes the other at colliders is best addressed. A
value of $\kappa_l\lapprox 0.2$ leads to a heavy neutrino LTP, whose
phenomenology is somewhat different from that of SUSY with a
neutralino LSP.

Thus $f$, together with $\kappa_{q,l}$, determines the part of the LHT
spectrum relevant for us. The mass of the triplet scalar $\Phi$ is
related to the doublet Higgs mass by
\be
m_\Phi = \sqrt{2} m_H {f \over v}. 
\ee
We will take $m_H = 120$~GeV throughout this paper. Two more dimensionless
parameters $\lambda_1$ and $\lambda_2$ appear in the top quark sector; the top
mass being given by $m_t = (\lambda_1 / \sqrt{1 + R^2}) v$ and $R = \lambda_1
/ \lambda_2$. The masses of the two heavy partners of the top quark, $T_{+}$
and $T_{-}$, can be expressed as
\be
m_{T_{+}} = \lambda_2 \sqrt{1 + R^2} f, \quad 
m_{T_{-}} = \lambda_2 f. 
\ee
We use $m_t = 171.4~\mbox{GeV}$ in our analysis and set $R = 1$, although this
does not have any significant bearing on our analysis.

\begin{figure}[t]
 \begin{center}
     \resizebox{120mm}{!}{\includegraphics{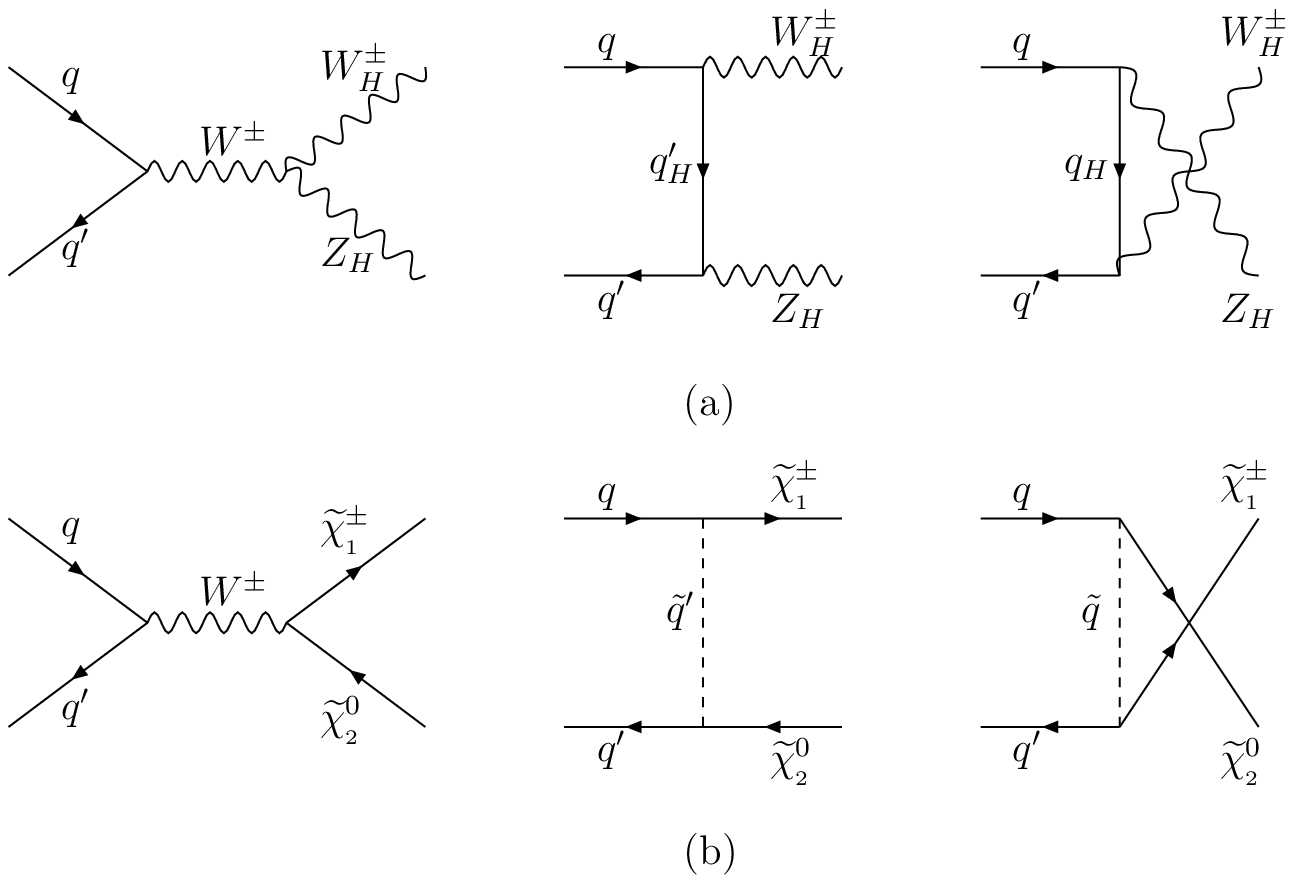}}
\caption{Representative leading order Feynman graphs contributing to the
pair production of $W_H^{\pm} Z_H$ in the LHT (a) and to $\chione \chitwo$ in
SUSY (b) at the LHC.}
\label{FG}
\end{center}
\end{figure}

Hadronically quiet final states comprising of trileptons can be produced in an
LHT scenario via $q\bar{q^{\prime}} \to W_H^{\pm} Z_H$ (see
Figure~\ref{FG}(a)). The most obvious way to trileptons from this is
$W_H^{\pm} \to A_H W^{\pm} \to A_H l^{\prime \pm} \nu_{l^{\prime}}$ and $Z_H
\to A_H Z \to A_H l^{\pm} l^{\mp}$.  However, the SM backgrounds, of which
WZ-production is the dominant source, need to be eliminated, the easiest way
being to disallow events with the invariant mass of any two opposite-sign
leptons in the neighbourhood of the $Z$-mass.  While this takes away many
signal events, one can still have trileptons if the T-odd heavy leptons are
lighter than the $Z_H$. In that case, the decay channel $Z_H \to {l_H^{\pm}}
l^{\mp}$ followed by ${l_H^{\pm}} \to A_H l^{\pm}$ opens up, and the trilepton
events can be quite copious in such a region of the parameter space.  As an
example, one can see that for $\kappa_l = 0.4$ and $f = 500~\mbox{GeV}$ one
has $m_{Z_H} =317$ GeV, $m_{{l_H^{\pm}}} = 283$ GeV and $m_{A_H} = 65$~GeV. We
shall comment later on the case where a real $l_H$ cannot be produced in
$Z_H$-decays.
%--------------------------------------------------------------------

Similar signals in SUSY are well-studied by now at hadron
colliders~\cite{Baer_et_al_Tevatron,Baer_et_al_LHC}; the main production
channel in a minimal model being $q\bar{q^{\prime}} \to \chione \chitwo$ (see
Figure~\ref{FG}(b)), with $\chione \to\nu_{l^{\prime}} {\widetilde l^{\prime
\pm}} \to \nu_{l^{\prime}} {l^{\prime \pm}} \chionezero$ and $\chitwo \to
l^{\pm} {\widetilde l^{\mp}} \to l^{\pm} l^{\mp} \chionezero $. {\em The
signals are not affected by the invariant mass cut so significantly as in the
case of LHT. This feature (namely, the susceptibility to invariant mass cut)
itself enables discrimination between the two scenarios.} However, as we shall
see below, clear quantitative distinction can be made from the predicted
strength of the signal as a whole.

The other major source of trileptons can be the production of heavy quarks
followed by their cascade decays into leptons via ${W_H^{\pm}}$ or $Z_H$
decays. The SUSY counterpart of such processes will be the production of
squarks followed by their cascade decays into leptons via chargino and
neutralino decays.  Such processes will be accompanied by two jets, but if
these jets are extremely soft, they can escape detection and therefore such
events can be misidentified as hadronically quiet. As we shall see later, our
jet recognition criteria disallow such final states, and thus one can fully
concentrate on final states which genuinely originate in the electroweak
sector. 

To see whether the LHT signal may be mimicked by the corresponding
supersymmetric signal, one has to go to situations where their particle
spectra have a close correspondence. It is assumed that the masses of the
particles produced in the hard scattering, and that of the invisible particle
(LTP/LSP) in the final state, can be extracted from various kinematic
distributions. These masses, it has been claimed, can be estimated at the LHC
up to an uncertainty of about $20-30$~GeV for SUSY
particles~\cite{SUSY_masses}. These references also indicate that the
uncertainty can be much less in situations where the masses are correlated, as
in a supergravity scenario. It is not unreasonable to expect a similar level
of precision in the case of the LHT where the masses of $W_H, Z_H$ and $A_H$
are all related to the parameter $f$.

At each chosen value of (f, $\kappa_q, \kappa_l$) in our analysis, we equate
the masses of the squarks and sleptons to those of the heavy quarks and
leptons determined by the relations given in
Eq.~(\ref{mirror_quark_masses}). Next, we try to align the heavy gauge boson
masses in Eq.~(\ref{gauge_boson_masses}) to those of the low-lying neutralinos
and charginos ($\chionezero, \chitwo, \chione$) which play a crucial role in
the production of trileptons. This is best done in a minimal SUSY scenario
where the gaugino masses ($M_1, M_2$) are not constrained by the requirement
of unification at a high scale.  $M_1$, the Bino mass, is set equal to
$m_{A_H}$. Next, we set a correspondence between ($m_{\chione},m_{\chitwo}$)
and ($m_{W_H}, m_{Z_H}$) for both the cases where the former pair is dominated
by the Wino and the Higgsino.  This is done by adopting two scenarios, namely,
(a) $M_1=m_{A_H}$, $M_2=m_{Z_H}$ and $\mu=1.5$ TeV (henceforth to be called
the SS1 scenario), and (b) $M_1=m_{A_H}$, $\mu=m_{Z_H}$ and $M_2=1.5$ TeV
(henceforth to be called the SS2 scenario). The physical chargino and
neutralino states are subsequently obtained by diagonalisation of the
respective mass matrices, and, as seen in Table~\ref{mass_table}, they indeed
demonstrate a very close resemblance of the spectra between LHT and
SUSY.\footnote{The states $\chione$ and $\chitwo$ should be close to each
other in mass (like $W_H$ and $Z_H$), but considerably heavier than
$\chionezero$ (like $A_H$). Matching the LHT spectrum, controlled by the
parameter $f$, in this manner is not possible with $\mu \lapprox
M_1$. Consequently, $\chionezero$ remains Bino-dominated in our analysis.}
%
%--------------------------------------------------------------------
\bt
\begin{center}
\begin{tabular}{|rr@{.}lr@{.}lr@{.}lr@{.}lr@{.}lr@{.}l|cr@{.}lr@{.}lr@{.}l|}
\hline \hline 
\multicolumn{13}{|c|}{\bf LHT} & \multicolumn{7}{c|}{\bf SUSY} \\ 
\hline \hline  
\multicolumn{1}{|c}{$f$} & \multicolumn{2}{c}{$m_{A_H}$} &
\multicolumn{2}{c}{$m_{Z_H}$} & \multicolumn{2}{c}{$m_{d_H}$} &
\multicolumn{2}{c}{$m_{u_H}$} & \multicolumn{2}{c}{$m_{l_H}$} &
\multicolumn{2}{c|}{$m_{\nu_H}$} & 
{\bf Case} & \multicolumn{2}{c}{$m_{\chionezero}$} &
\multicolumn{2}{c}{$m_{\chitwo}$} & \multicolumn{2}{c|}{$m_{\chione}$} \\  
\hline 
\multicolumn{13}{|c|}{$\kappa_l=\kappa_q=1$} & \multicolumn{7}{c|}{} \\ 
\hline
500 & 66 & 2 & 316 & 7 & 707 & 1 & 685 & 7 & 707 & 1 & 685 & 7 & 
{\bf SS1} & 65 & 9 & 314 & 9 & 314 & 9 \\ 
\multicolumn{13}{|c|}{} & {\bf SS2} & 63 & 7 & 314 & 9 & 318 & 1 \\ 
\hline 
1000 & 150 & 2 & 648 & 3 & 1414 & 2 & 1403 & 5 & 1414 & 2 & 1403 & 5 & 
{\bf SS1} & 149 & 8 & 645 & 0 & 645 & 0 \\ 
\multicolumn{13}{|c|}{} & {\bf SS2} & 148 & 9 & 645 & 0 & 646 & 2 \\ 
\hline \hline
\multicolumn{13}{|c|}{$\kappa_l=0.4$, ~ $\kappa_q=1$} & \multicolumn{7}{c|}{}
\\ 
\hline 
500 & 66 & 2 & 316 & 7 & 707 & 1 & 685 & 7 & 282 & 8 & 274 & 2 & 
{\bf SS1} & 65 & 9 & 314 & 9 & 314 & 9 \\ 
\multicolumn{13}{|c|}{} & {\bf SS2} & 63 & 7 & 314 & 9 & 318 & 1 \\ 
\hline 
1000 & 150 & 2 & 648 & 3 & 1414 & 2 & 1403 & 5 & 565 & 7 & 561 & 4 & 
{\bf SS1} & 149 & 8 & 645 & 0 & 645 & 0 \\ 
\multicolumn{13}{|c|}{} & {\bf SS2} & 148 & 9 & 645 & 6 & 646 & 0 \\ 
\hline \hline
\end{tabular} 
\caption{LHT and SUSY mass spectrum in GeV for fixed $f$, $\kappa_l$ and
$\kappa_q$.  The Higgs mass and top mass are taken to be $m_H=120$ GeV and
$m_t=171.4$ GeV, respectively.  SUSY masses are obtained by fixing
$M_1=m_{A_H}$, $M_2=m_{Z_H}$ and $\mu=1500$ GeV for scenario SS1, whereas for
SS2 $M_1=m_{A_H}$, $\mu=m_{Z_H}$ and $M_2=1500$ GeV. The other SUSY parameters
are $\tan\beta =10$ and $m_A=850$ GeV. Note that sfermion masses are taken to
be exactly equal to the respective mirror fermion masses and $M_3=5000$ GeV.}
\label{mass_table}
\end{center}
\et
%-----------------------------------------------------------------------

We assume $M_3$ to be $5$ TeV, thus decoupling the gluinos as desired.  For
the sake of simplicity, we set all the trilinear couplings ($A$) to zero,
except $A_t$, which we tune to get the lighter CP-even Higgs mass $m_H = 120$
GeV as in the case of the LHT. Our analysis is not affected by this tuning.
Note that it is not possible to match all particles in the minimal
supersymmetric standard model (MSSM) into corresponding states in the LHT and
{\it vice-versa}; for instance, the heavy quarks $T_{\pm}$ do not have a
counterpart in the MSSM. On the other hand, there are no states in the LHT
that would correspond to the heavier chargino and neutralinos. Furthermore,
the rest of the Higgs sector (the charged scalar, the heavier neutral scalar
and the pseudoscalar in SUSY, and the triplet states in LHT) does not
correspond similarly.  This, too, does not affect the signals under
consideration here.  Finally, we use $\tan\beta = 10$ and $m_A =
850~\mbox{GeV}$ throughout for the calculations in the MSSM.

We use the CalcHEP 2.5.i~\cite{CalcHEP} model file for the LHT written by the
authors of Ref.~\cite{Belyaev_et_al} to calculate cross-sections and branching
fractions. For the subsequent simulations, the cross-sections generated with
CalcHEP are then interfaced into PYTHIA 6.410~\cite{Pythia}. The
cross-sections and branching fractions for the MSSM are calculated directly in
PYTHIA. The parton densities for the calculation of cross-sections at the LHC
are evaluated at leading order using CTEQ6L~\cite{CTEQ} with renormalisation
and factorisation scale fixed by $\mu_R=\mu_F = \sqrt{\hat s}$.

%--------------------------------------------------------------------
 \begin{figure}[t]
    \begin{center}
      \resizebox{75mm}{!}{\includegraphics{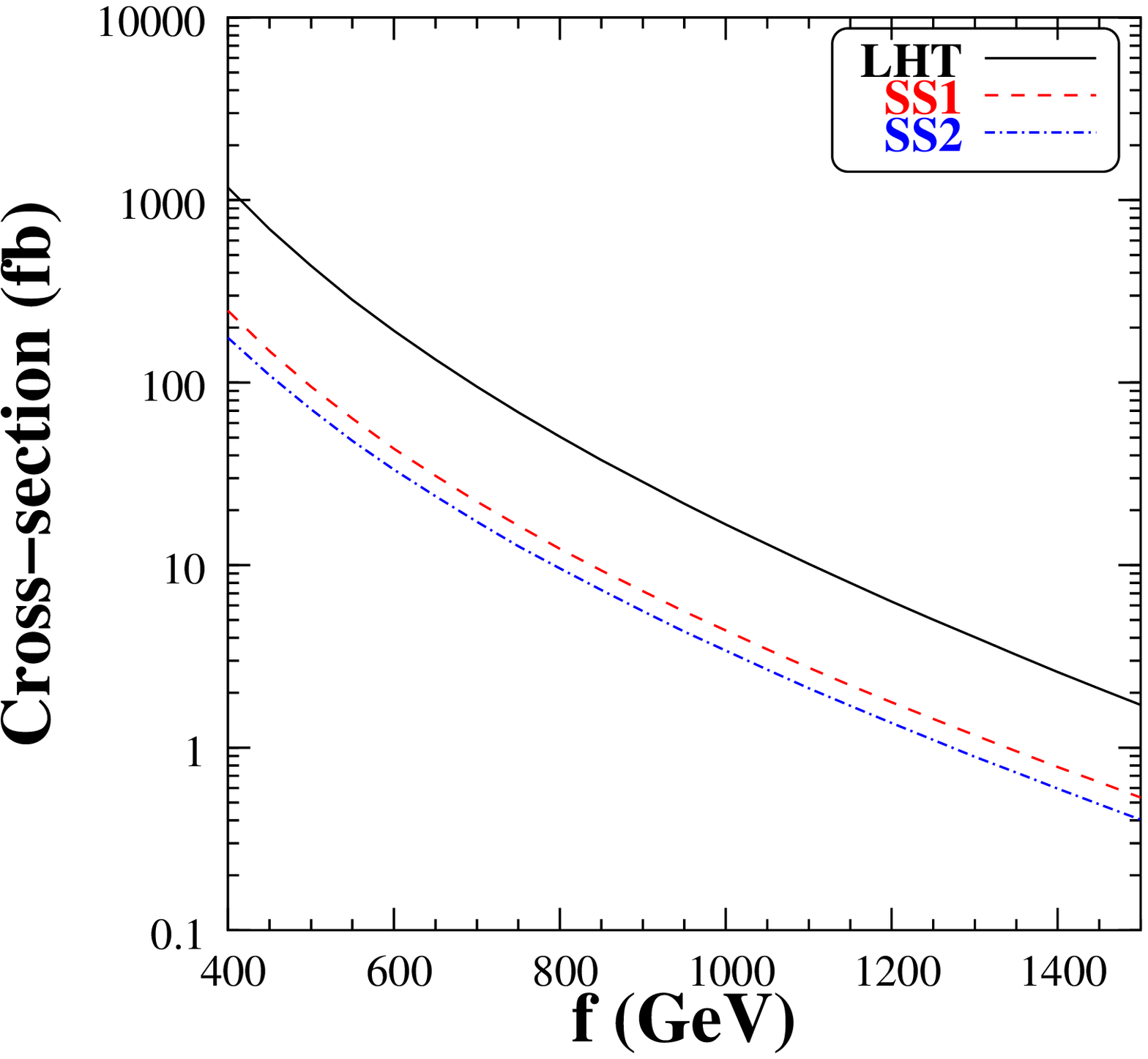}}
      \resizebox{75mm}{!}{\includegraphics{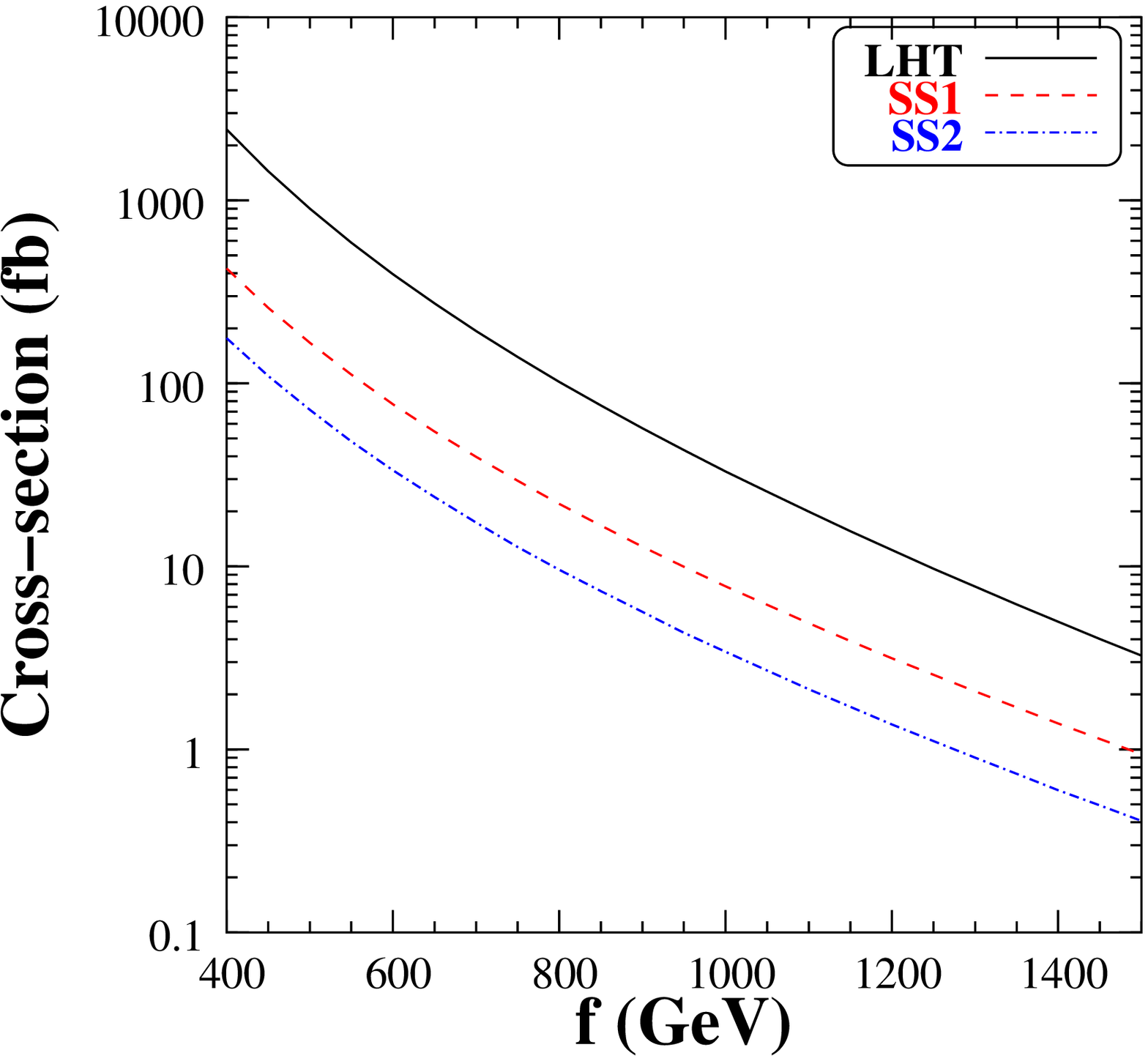}}
      \caption{Pair production cross-sections at the LHC of $W_H^{\pm}
	  Z_H$ (LHT) and $\chione \chitwo$ (SUSY) for $\kappa_q = 1$ (left
	  panel) and $\kappa_q = 1.5$ (right panel). The solid (black) line
	  represents the cross-section in the LHT.  The dashed (red) line
	  correspond to the SUSY scenario SS1 ($M_2 < \mu$) and the
	  dash-dotted (blue) line to the SUSY scenario SS2 ($\mu < M_2$). For
	  each value of $f$ the mass spectra in the two SUSY cases have been
	  matched to the LHT spectrum as described in the text.}
\label{sigma}
\end{center}
\end{figure}
%--------------------------------------------------------------------

In Figure~\ref{sigma} we plot the pair production cross-sections at the LHC
for $W_H^{\pm} Z_H$ with $\kappa_q=1$ and $\kappa_q=1.5$, as functions of the
LHT scale $f$.  As we vary $f$ and scan the LHT spectrum, the cross-sections
for $\chione \chitwo$-production have also been calculated (both for SS1 and
SS2), the SUSY spectrum being matched at each point as described above. The
difference between the LHT and the SS1 cross-section can be partially
attributed to the vector {\it vis-a-vis} fermionic final states in the
respective signals. A further suppression of the SS2 cross-section in
comparison to that of SS1 is also noticeable. This is because for SS2 the
couplings involved in the $t$- and $u$-channels are predominantly Yukawa in
nature and thus suppressed for light quarks from the proton beams, while for
SS1 the gauge coupling $g_2$ plays the vital role (see
Figure~\ref{FG}(b)). The cross-section enhancement with increased $\kappa_q$,
i.e. increased masses for the mirror quarks, shows that we are in a region of
the parameter space where the not so unusual destructive interference between
the $s$- and $t$-channel processes in Figure~\ref{FG}(a) becomes less
effective with increase in $\kappa_q$~\cite{Freitas_Wyler,Belyaev_et_al}. A
similar interference effect occurs for the SS1 case in
Figure~\ref{FG}(b). Therefore, the relative difference between SS1 and SS2
increases when going from lighter to heavier squarks when increasing
$\kappa_q$. In the case of SS2, with Higgsino dominated charginos and
neutralinos, it is mainly the $s$-channel diagram in Figure~\ref{FG}(b) which
contributes, so that there is very little difference between the
cross-sections for the two values of $\kappa_q$.

\begin{figure}
    \begin{center}
      \resizebox{55mm}{!}{\includegraphics{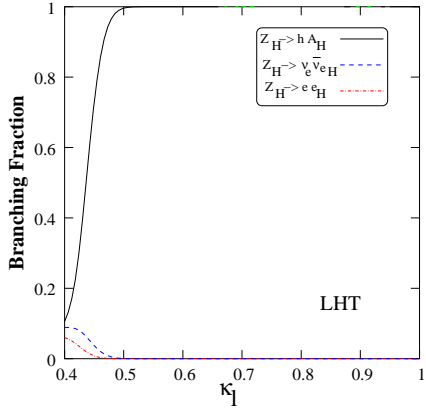}}
      \hspace*{2cm}\resizebox{55mm}{!}{\includegraphics{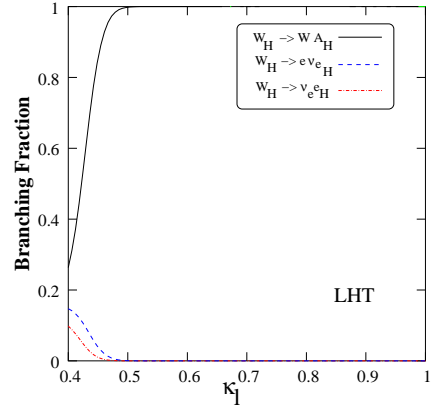}}
  \vspace*{-0.5cm}\begin{center} {\bf (a)}\end{center}
      \resizebox{55mm}{!}{\includegraphics{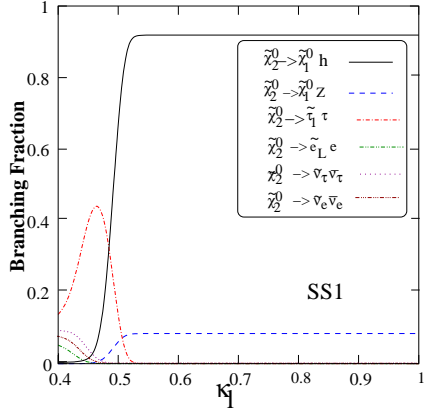}}
      \hspace*{2cm}\resizebox{55mm}{!}{\includegraphics{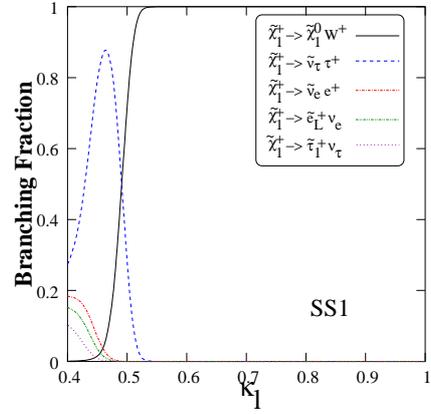}}
  \vspace*{-0.5cm}\begin{center} {\bf (b)}\end{center}
      \resizebox{55mm}{!}{\includegraphics{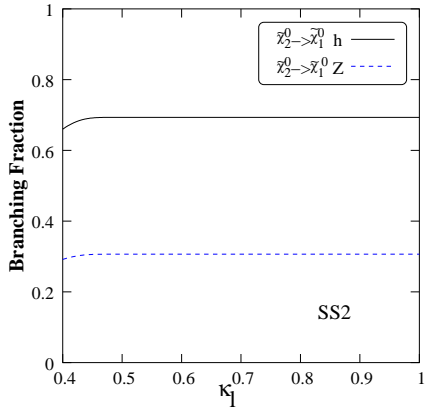}}
      \hspace*{2cm}\resizebox{55mm}{!}{\includegraphics{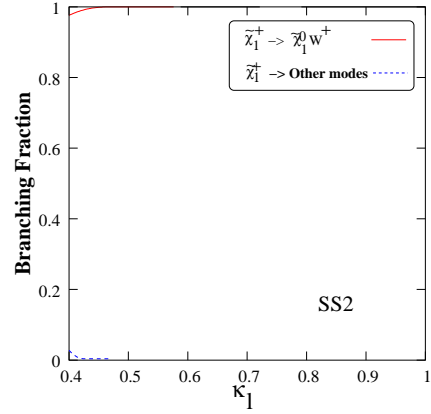}}
  \vspace*{-0.5cm}\begin{center} {\bf (c)}\end{center}
      \caption{LHT and SUSY branching fractions for the produced particles as
	  a function of $\kappa_l$ for fixed $f=500~{\rm GeV}$ and $\kappa_q =
	  1$. Panel (a) represents the LHT and panels (b) and (c) correspond
	  to SUSY scenario SS1 $(M_2 < \mu)$ and SS2 $(\mu < M_2)$,
	  respectively.}
      \label{bf}
    \end{center}
  \end{figure}

While the production cross-sections are controlled by the parameter $\kappa_q$
for a fixed $f$, the decay rates are primarily governed by $\kappa_l$, see
also Ref.~\cite{Cao_Chen}. In Figure~\ref{bf}(a-c) we plot the branching
fractions of the particles produced in the initial hard scattering for LHT, SS1
and SS2, respectively, as functions of $\kappa_l$, with the SUSY spectrum
appropriately matched. In Figure~\ref{bf}(a) (\ref{bf}(b)), we see that the
leptonic branching fractions are larger for $W_H^{\pm}~(\chione)$ or
$Z_H~(\chitwo)$ up to $\kappa_l=0.44$. This is because the masses of the heavy
leptons (sleptons) are smaller than the masses of the heavy gauge bosons
(chargino and neutralino). Above $\kappa=0.44$, the decays are purely into the
LTP (LSP) and a gauge boson or a Higgs. In case of SS1, as the produced
particles are gaugino dominated, their decays are governed by gauge couplings,
whereas for SS2 the produced particles being Higgsino dominated, it is the
Yukawa coupling which enters in the decay. This explains why the leptonic
branching fractions for SS1 (Figure~\ref{bf}(b)) are higher compared to SS2
(Figure~\ref{bf}(c)).
%
%--------------------------------------------------------------------
%--------------------------------------------------------------------

The event analysis is performed with PYTHIA at the parton level, turning off
initial- and final-state radiation.  To select our final trilepton states, we
apply the following cuts on our sample events:

\begin{itemize}
\item 
In order that the events are hadronically quiet, we reject jets having
${p_T}_j > 30~\mbox{GeV}$ and $|\eta_j|< 2.7$. This reduces the $t\bar{t}$
background considerably~\cite{Baer_et_al_LHC}.
\item 
Each lepton should have ${p_T}_l > 25$ GeV and $|{\eta}_l|\leq 2.5$, 
to ensure that they lie within the coverage of the detector.
\item 
$\Delta R_{ll} \geq 0.2$, (where $(\Delta R)^2 = (\Delta \eta)^2 +
(\Delta \phi)^2$) such that the leptons are well resolved in space.
\item 
A missing transverse energy cut, $E_T{\!\!\!\!\!\!/\ } \geq 100$ GeV 
has been employed to suppress the SM background.
\item 
We analyse only those events where $m_{l^+l^-} >20$ GeV which ensures the
absence of leptons emitted from off-shell photons.  An additional cut in the
form of $m_{l^+l^-} < m_Z -15$ GeV or $m_{l^+l^-} > m_Z +15$ GeV is used, in
order to eliminate the SM backgrounds from on-shell $Z$-bosons. Furthermore,
we demand $m_T{(l E_T{\!\!\!\!\!\!/\ })} <m_W -15$ GeV or $ m_T{(l
E_T{\!\!\!\!\!\!/\ })}>m_W +15$ GeV to reduce the backgrounds arising from
$W$-bosons~\cite{W_cut}.
\end{itemize}

%-----------------------------------------------------------------------
\bt
\begin{center}
\begin{tabular}{|c|c|r@{.}l|c|r@{.}l|}
\hline \hline
{\bf Cuts} & {\bf LHT} & \multicolumn{2}{|c|}{\bf SS1} & {\bf SS2} &
\multicolumn{2}{|c|}{\bf Background} \\ 
\hline \hline 
No jet with $p_{Tj}>30$ GeV and $|{\eta}_j|<2.7,$ & 
& \multicolumn{2}{|c|}{} & & \multicolumn{2}{|c|}{} \\ 
$p_{Tl}>25$ GeV, $|{\eta}_l|<2.5$ and $\Delta
R_{ll}>0.2$ & 9292.7 & 1641 & 4 & 68.1 & \hspace*{0.5cm}20232 & 5 \\ 
and $E_T{\!\!\!\!\!\!/\ }>30$ GeV & 
& \multicolumn{2}{|c|}{} & & \multicolumn{2}{|c|}{} \\ 
\hline 
$E_T{\!\!\!\!\!\!/\ } >100$ GeV & 7281.2 & 1187 & 6 & 49.6 & 1599 & 9 \\ 
$m_{l^\pm l^\mp}>20$ GeV & 7085.4 & 1137 & 5 & 48.1 & 1596 & 5 \\ 
$|m_{l^\pm l^\mp}-m_Z|>15$ GeV & 4543.9 & 659 & 8 & 18.2 & 467 & 1 \\ 
$|m_T{(l E_T{\!\!\!\!\!\!/\ })} -m_W|>15$ GeV & 4246.3 & 606 & 5 & 17.0 & 263
& 9 \\ 
\hline \hline     
\end{tabular}
\caption{Efficiency of the cuts on trilepton events at the LHC from $W_H^\pm
Z_H$ (LHT), $\chione\chitwo$ (SUSY) and from the SM background. The integrated
luminosity is assumed to be $300~{\rm fb}^{-1}$. The missing energy cut is
shown in two stages to convey the usefulness of the finally chosen value
$E_T{\!\!\!\!\!\!/\ } >100$ GeV. The values of the LHT parameters are $f=500$
GeV, $\kappa_q=1$ and $\kappa_l=0.4$. The SS1 and SS2 parameters corresponding
to this LHT point are given in Table~\ref{mass_table}.}
\label{cut_table}
\end{center}
\et
%-----------------------------------------------------------------------

\begin{figure}
    \begin{center}
      \resizebox{75mm}{!}{\includegraphics{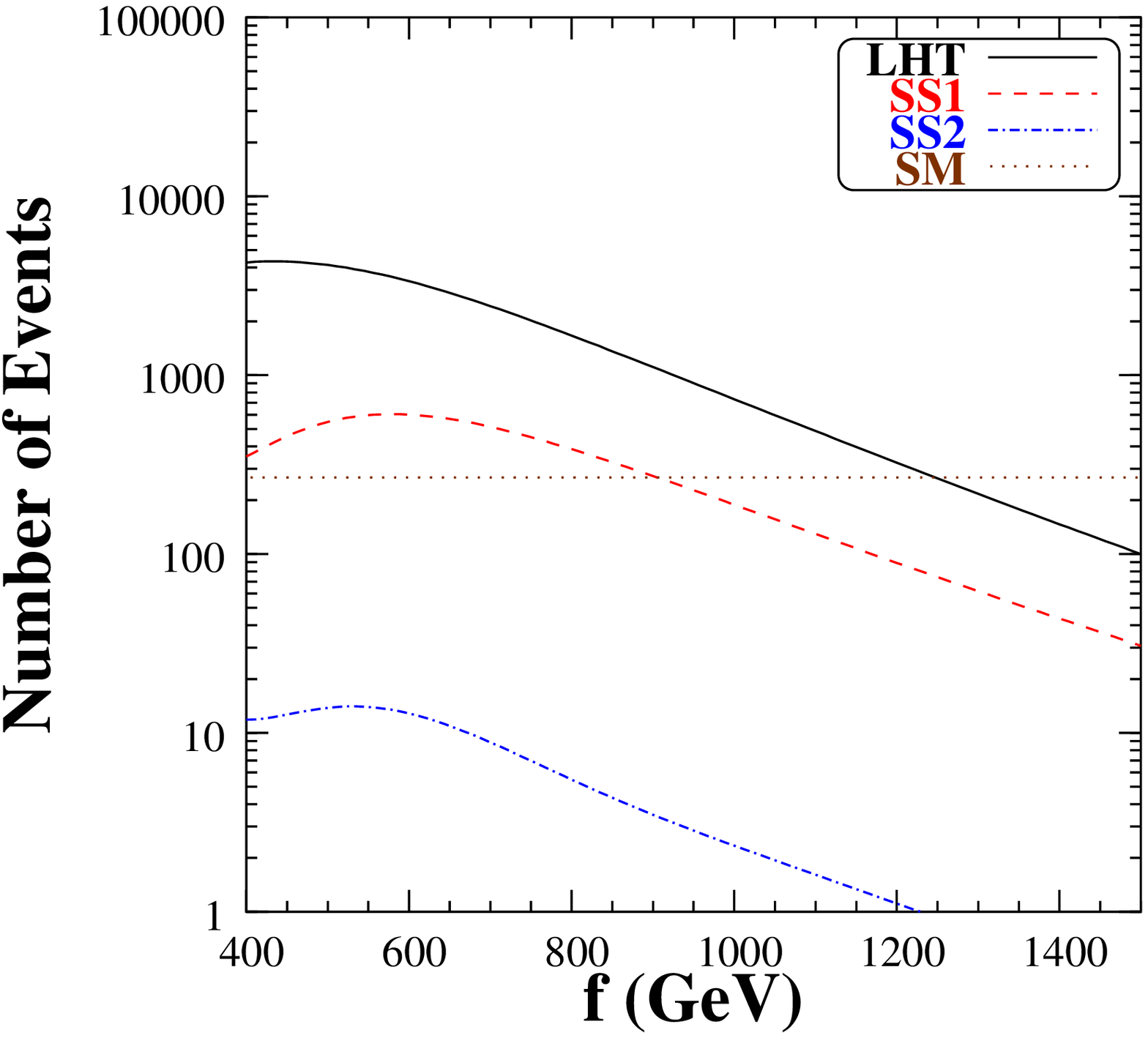}}
     \resizebox{75mm}{!}{\includegraphics{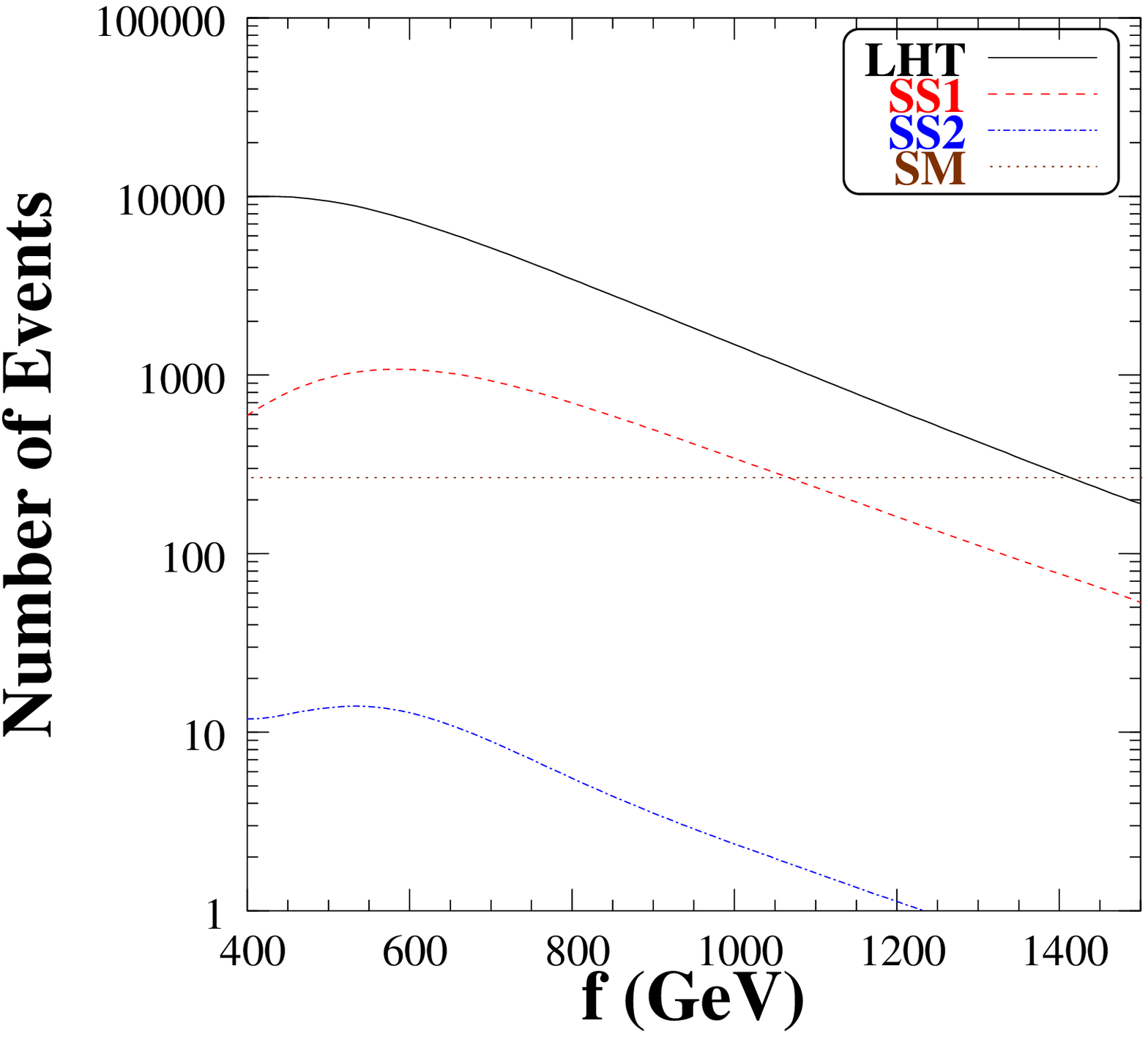}}
      \caption{Expected $3l + E_T{\!\!\!\!\!\!/\ }$ event rates after the cuts
	at the LHC from $W_H^\pm Z_H$ (LHT) and $\tilde \chi_1^\pm \tilde
	\chi_2^0$ (SUSY), for scenario SS1 ($M_2 <\mu$) and SS2 ($\mu < M_2$)
	with $\kappa_l=0.4$ for $\kappa_q = 1$ (left panel) and $\kappa_q =
	1.5$ (right panel). The integrated luminosity is assumed to be
	$300~{\rm fb}^{-1}$.  The line-style (colour) conventions for these
	plots are the same as in Figure~\ref{sigma}. The horizontal dotted
	line represents the SM background of 264 events after all the cuts.}
 \label{rates1}
    \end{center}
  \end{figure}

The efficiency of the cuts is shown in Table~\ref{cut_table}.  In
Figure~\ref{rates1} we present the variation of the number of trilepton events
against the scale $f$ for LHT, SS1 and SS2, after imposing the above
event-selection criteria. An integrated luminosity of $300~{\rm fb}^{-1}$ at
the LHC has been used for obtaining the number of events. This is done for
$\kappa_l = 0.4$, with $\kappa_q = 1.0$ and $\kappa_q = 1.5$, respectively. We
find that the LHT trilepton event rates remain higher after the cuts in
comparison to SS1 and SS2. This is primarily because of the larger
cross-sections for the LHT. The SS2 rates are further suppressed in comparison
to SS1 because of the small branching fractions for the leptonic decays of
$\chione$ and $\chitwo$.
%
%--------------------------------------------------------------------
%--------------------------------------------------------------------

As mentioned earlier, the production of heavy quarks (squarks), followed by
their cascade decays, might also lead to hadronically quiet trilepton events,
if the accompanying jets are very soft. We simulated such events and found
them to be negligible, since, with such masses as chosen here, the jets in the
final state almost always emerge with $p_T > 30$ GeV. Also, in the SUSY cases,
the cascade decays of the heavier charginos and neutralinos do not yield a
significant number of trileptons after the cuts are imposed.

The number of SM background events surviving the cuts is shown as a flat
dotted line in Figure~\ref{rates1}. A closer look at the figure shows that the
LHT trilepton events can be clearly distinguished, at least at the $6\sigma$
level, from either of SS1 or SS2, even in the presence of the SM background,
up to $f \simeq 1.5$ TeV for $\kappa_q = 1.0$, and $f \simeq 1.7$ TeV for
$\kappa_q = 1.5$. Figure~\ref{rates1} also shows that a comparable number of
events for LHT and SS1 may only result from widely different values of $f$
(differing by almost 300~GeV or so). Thus, conservatively, the event rates are
likely to be still distinguishable if the uncertainty in $f$ is around 200~GeV
or less.  Actually even with an integrated luminosity of $30~{\rm fb}^{-1}$ a
differentiation between LHT and SUSY above the SM background could well be
possible, although of course with less significance. It should be noted,
however, that we did not take into account systematic errors such as
uncertainties from higher order QCD corrections, parton distributions,
initial- and final-state radiation and detector effects. Only a more detailed
and realistic analysis could show whether a distinction between the different
models can be made with lower statistics. Moreover, distinguishing between LHT
and MSSM will only be feasible, if we already have some knowledge about the
mass spectrum in the underlying model. Maybe not enough information on all the
masses will be available after the early phase of LHC with $30~{\rm fb}^{-1}$
of data. Once we have precise enough information on the relevant masses, LHT
and SS1 would be clearly distinguishable by the trilepton yield which, as is
clear from Figure~\ref{rates1}, would be an order of magnitude higher for LHT
than that for SS1 for similar masses.  The SS2 events, on the other hand, are
going to be well below the backgrounds, for the mass range under investigation
here.

A similar study for a higher $\kappa_l$ value, namely, $\kappa_l=1$ does not
yield fruitful results, as in this case the heavy leptons are more massive
than the heavy gauge bosons.  The only possible decay modes are ${W_H^{\pm}}
\to W^{\pm} A_H$ and $Z_H \to h A_H$. Thus the Higgs boson controls the number
of trilepton events in the higher $\kappa_l$ regions. In the LHT, below about
$f=470$ GeV, a Higgs boson with a mass of $120$ GeV decays invisibly into two
heavy photons about 90\% of the times~\cite{Hundi_Mukhopadhyaya_Nyffeler},
while beyond $f=470$ GeV, it decays into $b\bar{b}$ with a branching fraction
of about 70\%. Therefore, there are only a few trilepton events generated via
$h \to \tau\bar{\tau} \to l\bar{\nu}\nu_\tau \bar{l}\nu_l\bar{\nu}_\tau$.  In
fact, from Figure~\ref{bf}, it is clear that the above conclusions will remain
unaffected for any $\kappa_l$ beyond $\kappa_l>0.44$, because the mirror
leptons remain heavier than the heavy gauge bosons in this $\kappa_l$
region. The trilepton signals from neither LHT nor SUSY (with its slepton
masses correspondingly higher) can rise above the SM backgrounds in this
region.
 
Earlier studies have predicted appreciable rates for trilepton signals
in SUSY at the Tevatron~\cite{Baer_et_al_Tevatron} and the
LHC~\cite{Baer_et_al_LHC}.  However, they analysed regions with lower
masses than what has been considered here.  Since the values of $f$ in
the LHT corresponding to such masses are ruled out by precision
electroweak observables, such regions are not pertinent to the
distinction between the LHT and SUSY scenario. Moreover, the choice of
cuts in these studies is different from ours.

Before we end, we want to reiterate that, apart from ensuring
$m_{\nu_H}>m_{A_H}$, we have not made any 'tailored' parameter
choice. The masses of SUSY particles are kept at par with those of the
LHT spectrum in each case, in order to have similar event
kinematics. Also, both the cases of gaugino and Higgsino domination in
the lighter chargino and the second lightest neutralino are included
in our study, making the comparisons practically exhaustive.

In conclusion, we have analysed hadronically quiet trilepton events
arising in both T-parity conserving Littlest Higgs and R-parity
conserving SUSY models at the LHC. We found a clear excess of
trilepton events in the LHT over the corresponding number of events in
the two SUSY scenarios with a mass spectrum that matches the one in
the LHT. While for $\kappa_l \le 0.44$, it is possible to rise above
the Standard Model backgrounds, such backgrounds become a problem for
larger values of $\kappa_l$. Therefore, while the hadronically quiet
trilepton signal suggests a promising way of distinguishing between
SUSY and LHT, this signal is perhaps best usable if the heavy leptons
(sleptons) do not exceed the heavy gauge bosons ($\chitwo/\chione$) in
mass.

%------------------------------------------------------------------------

\vspace*{0.4cm}
\noindent
{\bf Acknowledgments} \\ 
This work was partially supported by the Department of Atomic Energy,
Government of India, under a 5-Years Plan Project. Computational work for this
study was carried out at the cluster computing facility in the Harish-Chandra
Research Institute (http:/$\!$/cluster.mri.ernet.in).

%------------------------------------------------------------------------

\end{document}